\documentclass{article}

\usepackage{PRIMEarxiv}

\usepackage[utf8]{inputenc} % allow utf-8 input
\usepackage[T1]{fontenc}    % use 8-bit T1 fonts
\usepackage{hyperref}       % hyperlinks
\usepackage{url}            % simple URL typesetting
\usepackage{booktabs}       % professional-quality tables
\usepackage{amsfonts}       % blackboard math symbols
\usepackage{nicefrac}       % compact symbols for 1/2, etc.
\usepackage{microtype}      % microtypography
\usepackage{lipsum}
\usepackage{fancyhdr}       % header
\usepackage{graphicx}       % graphics
\graphicspath{{media/}}     % organize your images and other figures under media/ folder
\usepackage{braket}
\usepackage{mathtools}
\usepackage{hhline}
\usepackage{tikz}
\usetikzlibrary{calc}
\usepackage{authblk}
\usepackage{subcaption}

\usepackage{changes}

\title{Solving The Quantum Many-Body Hamiltonian Learning Problem with Neural Differential Equations}

\author[1,2]{Timothy Heightman\thanks{Corresponding author: theightman@icfo.eu}
}
\author[1]{Edward Jiang}
\author[1,3]{Antonio Acín}
\affil[1]{\textit{ICFO-Institut  de  Ciencies  Fotoniques,  The  Barcelona  Institute  of  Science  and  Technology, 08860 Castelldefels (Barcelona), Spain}}
\affil[2]{\textit{Quside Technologies SL, Carrer d’Esteve Terradas, 1, 08860 Castelldefels, Barcelona, Spain}}
\affil[3]{\textit{ICREA-Institucio Catalana de Recerca i Estudis Avan\c cats, Lluis Companys 23, 08010 Barcelona, Spain}}

\begin{document}
\maketitle

\begin{abstract}
Understanding and characterising quantum many-body dynamics remains a significant challenge due to both the exponential complexity required to represent quantum many-body Hamiltonians, and the need to accurately track states in time under the action of such Hamiltonians. This inherent complexity limits our ability to characterise quantum many-body systems, highlighting the need for innovative approaches to unlock their full potential. To address this challenge, we propose a novel method to solve the Hamiltonian Learning (HL) problem—inferring quantum dynamics from many-body state trajectories—using Neural Differential Equations combined with an Ansatz Hamiltonian. Our method is reliably convergent, experimentally friendly, and interpretable, making it a stable solution for HL on a set of Hamiltonians previously unlearnable in the literature. In addition to this, we propose a new quantitative benchmark based on power laws, which can objectively compare the reliability and generalisation capabilities of any two HL algorithms. Finally, we benchmark our method against state-of-the-art HL algorithms with a 1D spin-$1/2$ chain proof of concept.
\end{abstract}


\section{Introduction}
\label{sec:introduction}
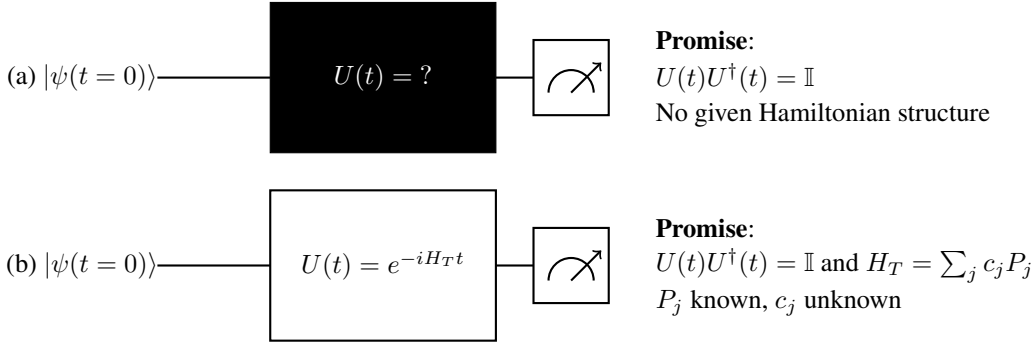
\begin{figure}
    \centering
\begin{tikzpicture}
    % Draw the black box and wires
    \draw[fill=black] (2,0) rectangle (5,2);
    \draw[thick] (0.5,1) -- (2,1);
    \draw[thick] (5,1) -- (5.5,1);
    \node at (-0.5,1) {(a) \(\ket{\psi(t = 0)}\)};
    
    \draw[thick] (5.5,0.5) rectangle (6.5,1.5);
    \draw[thick] (6.3,0.8) arc[start angle=0,end angle=180,radius=0.3];
    \draw[->][thick] (6.0,0.8) -- (6.4,1.2);
    
    % Draw question mark inside the black box
    \node[white] at (3.5,1) {\(U(t) = \;?\)};
    %\node[above] at (3.5, 2.3) {};
    \node[right] at (7.0,1.5) {\textbf{Promise}:};
    \node[right] at (7.0,1.0) {\(U(t) U^{\dagger}(t) = \mathbb{I}\)};
    \node[right] at (7.0,0.5) {No given Hamiltonian structure};

    % Draw white box and wires
    \draw[thick] (2,-2.5) rectangle (5,-0.5);
    \node at (3.5,-1.5) {\( U(t) = e^{-i H_T t} \)};
    \draw[thick] (0.5,-1.5) -- (2,-1.5);
    \draw[thick] (5,-1.5) -- (5.5,-1.5);
    \node at (-0.5,-1.5) {(b) \(\ket{\psi(t = 0)}\)};
    
    \draw[thick] (5.5,-1.0) rectangle (6.5,-2.0);
    \draw[thick] (6.3,-1.7) arc[start angle=0,end angle=180,radius=0.3];
    \draw[->][thick] (6.0,-1.7) -- (6.4,-1.3);

    % Add Hamiltonian structure known note to the right of the measurement operator
    \node[right] at (7.0,-1.0) {\textbf{Promise}:};
    \node[right] at (7.0,-1.5) { \(U(t) U^{\dagger}(t) = \mathbb{I}\) and \( H_T = \sum_j c_j P_j \)};
    \node[right] at (7.0,-2.0) {\(P_j\) known, \( c_j \) unknown};
    
\end{tikzpicture}
    \caption{The black-box and white-box Hamiltonian Learning (HL) scenarios are shown in (a) and (b) respectively. In both cases, the evolution is unitary, and we may control the length of time, $t$, for which the system evolves. In the white-box scenario (a), the structure of the true Hamiltonian \( H_T = \sum_j c_j P_j \) is known, but the coefficients \( c_j \in \mathbb{R} \) are not. Whilst in the black-box scenario (b), neither the structure of the Hamiltonian, $P_j$, nor the coefficients $c_j$ are known. As discussed in Sec. \ref{sec:introduction}, we choose to formulate the HL problem assuming control on the amount of time evolution, as it requires little optimal control. This makes data from time-evolved states experimentally friendly, provided the chosen evolution times are sufficiently short.}
    \label{fig:white_box_black_box}
\end{figure}

Characterizing many-body Hamiltonians is central to the development of quantum information and processing technologies. Once the true Hamiltonian generating unitary dynamics is known, we can use this knowledge to perform error mitigation, optimal quantum control, quantum simulation, and device certification \cite{strikis2021learning, bennewitz2022neural, xie2022bayesian}. The Hamiltonian Learning (HL) problem is defined as the task of inferring the Hamiltonian of a many-body system given some dataset about that system. In the context of quantum many-body systems and information processing, this task is difficult for two key reasons. First, there is an exponential number of interactions that a many-body system can have, making the search space exponentially large. Second, there are sets of Hamiltonians that can be compatible with a data-set taken from a quantum system; the answer may not be unique.  This gives rise to two equally important formulations of this problem shown in Fig.~\ref{fig:white_box_black_box}, hereby referred to as the white-box and black-box scenarios. In the white-box scenario, we assume that the \textit{structure} of the Hamiltonian is known, and inferring it simply involves tuning the interaction coefficients to best match the dataset. This means the set of compatible Hamiltonians is relatively small, even being unique in some cases \cite{hou2020determining,qi2019determining}. In contrast, the black-box scenario makes no such assumptions. Here, the structure of the Hamiltonian must also be found by representing it as a sum over different quantum operators with weighted interaction coefficients. Consequently, the set of Hamiltonians that are compatible with the data is larger, comprising of different possible operators, as well as their relative weighting. \\

There are a variety of approaches to try and tackle these two versions of the HL problem \cite{wang2022quantum, wiebe2014quantum, gebhart2023learning, shi2022parameterized, gu2022practical, wilde2022scalably, dutkiewicz2023advantage, hou2020determining, bairey2019learning, qi2019determining}. By exploring which types of quantum states can be informative, theoretical studies have demonstrated the conditions required to create a dataset from which a Hamiltonian can be efficiently learned. For example, works such as \cite{hou2020determining,qi2019determining} show how one can determine Hamiltonians from eigenstates, whereas \cite{bairey2019learning,anshu2021sample, haah2022optimal, bairey2020learning} address the use of thermal states to generate data. In contrast,  \cite{wilde2022scalably,qi2019determining, huang2023learning, huang2022learning, zhang2014quantum} show that using a selection of states and evolving them for a controlled amount of time can also be meaningful. Intuitively this can be considered as using the \textit{trajectories} of a few quantum states in order to find the most likely Hamiltonian. This idea is advantageous in practical settings as it involves preparing few quantum states, so there is less overhead in terms of state preparation when compared to other methods. 
Furthermore, evolving states under the action of the true Hamiltonian for variable amounts of time requires no optimal control \cite{dutkiewicz2023advantage}. We need only choose the time at which to measure the output states. For these reasons, we will focus on the task of inferring Hamiltonians from time-evolved states shown in Fig.~\ref{fig:white_box_black_box}. \\

Despite these considerations, HL algorithms are yet to meet the needs of theoreticians, experimentalists, and industrial hardware providers. From the theoretical perspective of quantum simulation, state-of-the-art HL algorithms are so far incapable of learning sufficiently complex Hamiltonians. This means that the space of interactions from which we can infer a meaningful representation is limited when attempting to simulate. For this limited set of interactions, we can often simulate through simpler means than HL. Whilst on the experimental side, state-of-the-art HL algorithms cannot handle noisy many-body systems robustly. This limits their applicability in optimal control and error mitigation \cite{valenti2019hamiltonian}, where the imperfections arising from unforeseen interactions in the system and environment must be taken into account to meaningfully perform quantum control and correct for systematic errors. Where they are robust to noise, current algorithms are rarely scalable to the level needed in current NISQ hardware \cite{decross2024computational, morgado2021quantum, kliesch2021theory}. This is either because they work on too few spins or do not cover a sufficiently large interaction set to fully characterise and certify NISQ hardware.

Recent advances in deep learning for quantum dynamical systems highlight the potential for its use to solve the HL problem \cite{schuld2021machine,carleo2019machine, dawid2022modern, marquardt2021machine, khait2022optimal, shi2022parameterized, gu2022practical}. These advances have provided useful tools in special cases like quantum state tomography \cite{torlai2018neural, bennewitz2022neural} and variational Monte-Carlo \cite{bennewitz2022neural, becca2017quantum}, gaining some traction in the wider quantum community \cite{schuld2021machine, dawid2022modern}. The philosophy behind recently developed deep-learning based methods is to use neural networks as function approximators.
For example, using neural networks to detect phase transitions posits the existence of a map from measurement data to probabilities of being in different phases \cite{dong2019machine}. Even when neural networks are used to represent quantum states, we rely on the existence of a function that maps configurations to probability amplitudes \cite{carleo2017solving}. However, a general deep learning framework is yet to solve the above pressing issues in HL, because in HL the fundamental object we are aiming to represent is not a function, but an \textit{operator}. Whilst some progress has been made using so-called neural quantum states formalism to represent density operators \cite{hartmann2019neural}, one cannot use such constructions to solve the HL problem. In simple terms, this is because our solution should facilitate unitary dynamics for any input quantum state (in principle), meaning an exponential number of forward passes would be needed to enumerate how every component of the basis evolves. Furthermore, the purpose of solving the HL problem is to return an estimator for the ground truth in an interpretable form. A neural network representation does not yet allow for such an interpretation.
% \toni{I WAS WONDERING ABOUT THIS AND HOW IT COMPARES TO YOUR PREVIOUS POINT ON STATES. CAN A NASTY REFEREE ARGUE THAT IN THIS CASE IS ALSO A FUNCTION TO THE SET OF COEFFICIENTS $c_j$?} 
In this sense, our aim is to leverage deep learning techniques to construct propagators of quantum many-body systems in a way that facilitates an intepretable answer. Some recent works have explored this concept in the context where the potential (and therefore the Hamiltonian) is known \cite{secor2021artificial, ullah2021speeding, ullah2022predicting, nikhil2018convolutional}. Their objective in this case is to predict the time evolution of states using neural networks and a \textit{given} action, which is not available in our problem setting as it must be inferred. This means techniques such as Kernel Ridge Regression \cite{ullah2021speeding}, non-recursive Artificial Intelligence-based Quantum Dynamics \cite{ullah2022predicting}, and One-Shot Trajectory Learning \cite{nikhil2018convolutional} cannot be used to solve the HL problem, where the action is not known a-priori in both scenarios of Fig.\ref{fig:white_box_black_box}.\\

In this work, we introduce a general framework to solve the HL problem using the language of Neural Differential Equations (NDE). A NDE uses a neural network to \textit{specify} a system of differential equations. This sits in contrast to typical uses of neural networks in the physical sciences, where a system of differential equations is given, and a neural network is used to numerically solve that system \cite{ullah2021speeding, nikhil2018convolutional,ullah2022predicting}. Our method presents several advantages in terms of the complexity of true Hamiltonians that are now learnable in our framework. It extends both the set of Hamiltonians now learnable, and offers accurate estimates of its generalisation capabilities. We demonstrate these advantages by using relatively small neural networks to benchmark our HL algorithm on 1D spin-$1/2$ chains for systems up to 8 spins.

We also propose a new benchmark for HL which is agnostic to the specifics of a given implementation. In brief, our benchmark explores the power-law scaling behaviour of a HL algorithm outside its training data, allowing us to estimate the quality and reliability of a HL algorithm independently of the Hamiltonians considered or the number of bodies. Using this benchmark amongst others found across the literature, we demonstrate the capabilities of our framework to comfortably surpass the state-of-the-art in terms of robustness, accuracy and generalisation capabilities. We then explain how our method is able to learn complex Hamiltonians by analysing loss landscapes, using our numerical study of 1D spin-$1/2$ chains as an instructive example. Finally, we highlight the interpretability of our framework, using curriculum learning to allow users to yield an interpretable estimate for the Hamiltonian, rather than a black-box neural representation, as depicted in Fig. \ref{fig:white_box_black_box}. Consequently, our method can be used in both the black-box scenario as done in \cite{gu2022practical}, as well as the white-box scenario of Fig. \ref{fig:white_box_black_box} \cite{wilde2022scalably}. This opens up possibilities to use our framework for error mitigation, quantum optimal control, quantum simulation, and device certification \cite{kliesch2021theory, wiebe2014hamiltonian, wang2017experimental}. \\

The rest of this work is organised as follows: In section \ref{sec:statement_of_problem} we formally define the HL problem. In section \ref{sec:background} we describe in detail how to use NDEs in quantum dynamics to solve the HL problem. We then recall how to construct loss functions which are experimentally friendly, as was done in \cite{wilde2022scalably}, and provide a curriculum learning procedure to yield an interpretable solution to the HL problem. In section \ref{sec:results} we probe the performance of our HL framework on representative problems from 1D spin-$1/2$ chains, showing how the robustness, accuracy and experimental friendliness surpass current state of the art methods such as \cite{gu2022practical,wilde2022scalably}. Finally, in section \ref{sec:conclusion} we discuss the implications of our findings in error mitigation, optimal quantum control, quantum simulation and device certification.

\subsection{Statement of Problem}
\label{sec:statement_of_problem}
For spin-$1/2$ systems such as qubits, we consider the unitary dynamics of an initial state, $\ket{\psi(t = 0)}$, governed by the Time Dependent Schrödinger Equation (TDSE),
\begin{equation}
    i \ket{\dot{\psi}} = H_T \ket{\psi},
    \label{eq:TDSE}
\end{equation}
where $H_{T}$ is the true Hamiltonian for the system's evolution. The HL problem aims to find a representation of $H_{T}$ that generates the correct output for given input states as shown in Fig.\ref{fig:white_box_black_box}. In general, we wish for the learned representation to be valid over arbitrary time-scales. However, in practise such a representation is usually valid for transient (short) times \cite{zhang2021learning}, or the steady state \cite{bairey2019learning, zhou2024recovery, bairey2020learning}. In this manuscript, we will work with 1D, spin-$1/2$ chains of $N$ spins, whose Hamiltonian may always be expressed in the Pauli basis, 
\begin{equation}
    H_T = \sum_{j = 0}^{4^N - 1}c_j P_j,
    \label{eq:H_T_decomp}
\end{equation}
where $P_j \in \mathcal{P}_N$ is an element of the Pauli group on $N$ spins, and $c_j \in \mathbb{R}$ are real coefficients. However, we emphasise that the philosophy behind our method goes beyond 1D spin-$1/2$ chains. We use them as an example for conceptual clarity and numerical studies to benchmark against other methods, who have broadly done HL on 1D spin-$1/2$ chains \cite{bairey2019learning, wilde2022scalably,bairey2020learning, qi2019determining, hou2020determining, huang2023learning, zhang2014quantum, che2021learning}.

Since there are an exponential number of learnable parameters over the whole Pauli group, many works make the HL problem more tractable by aiming to learn subgroups of $\mathcal{P}_N$, formed with assumptions such as random sparse support \cite{yu2023robust}, maximal $k$-Locality\footnote{often, geometric $k$-Locality is assumed, see \cite{usui2024simplifying, wu2023variational, ranard2021aspects} for further discussion.} \cite{usui2024simplifying, ranard2021aspects}, order of non-identity terms \cite{kempe2005complexity, wilde2022scalably, bairey2019learning} , or homogeneity of the coefficients $c_j$ \cite{granade2012robust, wilde2022scalably}. Such assumptions limit the complexity of the true Hamiltonian, allowing previous HL methods to work by reducing the number of free parameters that need to be found. However,
these assumptions impose theoretical limits on the effectiveness of HL, since they will be unable to accurately learn true Hamiltonians outside this subgroup. Intuitively, with too few parameters to tune in an Ansatz, we cannot explore a space of Hamiltonians that is large enough to cover all the possible interactions that may be happening in the physical system. It is therefore of interest to find a procedure which works over as large a set of interactions as possible. Our procedure makes no assumptions on the order of Pauli operators, nor the homogeneity of the coefficients $c_j$, thus avoiding the limiting effects on expressivity shown by previous methods. \\

It is crucial to be able to infer this Hamiltonian efficiently from the data, i.e. requiring sub-exponential resources to accurately characterise the quantum many-body system of interest. Whilst some procedures assume unbridled access to the output state or true Hamiltonian \cite{che2021learning}, it has been shown that there are many cases where we can learn a Hamiltonian from local measurements \cite{hou2020determining}. This means we can avoid cumbersome procedures like quantum state and process tomography \cite{mohseni2008quantum, anshu2024survey, torlai2023quantum, neugebauer2020neural} to construct estimators for the difference between the ground truth and our estimates. In line to previous state-of-the-art works such as \cite{wilde2022scalably}, we use an experimentally friendly format for training data which involves sampling from the marginal distributions of a simulated output state. As such, the loss function is constructed from binary bitstrings sampled from marginal distributions corresponding to compatible observable, as might be done on idealised experimental hardware. See Sec. \ref{sec:dataset} for further discussion and examples.

\section{Neural Differential Equations in Quantum Dynamics}
\label{sec:background}

A common practise in machine learning is to use deep neural networks as function approximators to solve a given system of differential equations \cite{raissi2019physics, mishra2023estimates, bertschinger2024training, elbrachter2021deep}. Whereas the central idea of Neural Ordinary Differential Equations (NODEs) is that we use neural networks to \textit{specify} differential equations \cite{chen2018neural, kidger2022neural}. For example, let $x(t) \in \mathbb{R}^N$, then we may define
\begin{equation}
    \frac{dx}{dt}(t) = f_{\theta}(x(t)),\qquad\qquad x(t = 0)=x_0,\label{eq:ode}
\end{equation}
where $f_{\theta}: \mathbb{R}^N \rightarrow \mathbb{R}^N$, is some neural architecture with trainable parameters $\theta$. The input to the NODE is the initial vector $x(t = 0) = x_0$, and the output is $x(T)$ which is obtained by numerical integration of equation (\ref{eq:ode}) up to some time $T$. Eq.(\ref{eq:ode}) exploits the Universal Approximation Theorem \cite{csaji2001approximation, lu2020universal} to predict the rate of change of a system of differential equations. \\

The NODE can be used as a layer embedded in some larger neural network. As such, the solution, $y(t),\; t \in [0,T]$, forms a differentiable computation graph. We may tune this computational graph by varying its parameters, $\theta$, until the numerical solver outputs a trajectory, $y(t)$, that matches some desired behaviour $y_{\text{true}}(t)$. One pertinent example is tuning with gradient descent,
\begin{equation}
\theta \gets\theta-\varepsilon\nabla_\theta\mathcal{L}(\theta),
\label{eq:grad_desc}
\end{equation}
where $\mathcal{L}:\mathbb{R}^N \times \mathbb{R}^N \rightarrow \mathbb{R}$ is any loss function mapping the tuple $(y(t),y_{\text{true}(t)})$ to the reals. However, in principle, any gradient-based optimisation strategy can work, for example we use the so-called ADAM optimiser \cite{kingma2014adam}. \\

For a comprehensive introduction to NODEs and their applications in dynamical systems, we refer the reader to \cite{kidger2022neural} and references therein. To integrate Eq.(\ref{eq:ode}), the choice of numerical solver is somewhat arbitrary, and comes with the usual trade-off between accuracy and speed\footnote{In this manuscript we use the \texttt{diffrax} package \cite{kidger2022neural} which integrates seamlessly with the Python-based numerical computation library, \texttt{jax}, opting for the third-order Runge Kutta method \cite{butcher1996runge} for our numerical integration. See Sec. \ref{sec:results} for further discussion on our implementation and numerical study.}. From the perspective of HL, we can consider trying to learn a representation of the true Hamiltonian, $H_T$ from Eqs.(\ref{eq:TDSE},\ref{eq:H_T_decomp}) by specifying an Ansatz structure, $H_A(\theta)$, and numerically integrating the TDSE, with the set of parameters, $\theta$, being tuned to best match observed trajectories. This formulation of parameter tuning after numerical integration can put the work of \cite{wilde2022scalably, gu2022practical, bairey2020learning} into the language of NDEs. For example, consider using as an Ansatz the Transverse Field Ising Model (TFIM) Hamiltonian,
\begin{equation}
    H_{\text{TFIM}} = \sum_{\braket{i,j}} J_{ij} X_i X_j + \sum_k h_k Z_k,
\end{equation}
where $X_j$ is the Pauli $x$-matrix on site $j$, $Z_k$ is the Pauli $z$-matrix on site $k$, $h_k$ is a local transverse field, and $\braket{i,j}$ specifies that only nearest neighbour sites are coupled. For this Ansatz, the tunable parameter set $\theta$ is $ = \{J_{ij}, h_k\} \in \mathbb{R}$. By comparing the output states found by numerically integrating
\begin{equation}
 \ket{\psi(t + \Delta t);\theta} = 
 \int_t^{t + \Delta t} \frac{d }{dt'} \ket{\psi(t')}dt' = 
 -i \int_t^{t + \Delta t} H_{\text{TFIM}}(\theta) \ket{\psi(t')}dt',
\end{equation}
with true output states, 
\begin{equation}
    \ket{\psi(t)}_{\text{true}} = e^{-i H_T t} \ket{\psi(t = 0)}
\end{equation}
we can construct a simple Loss function, $\mathcal{L}:\mathcal{H} \times \mathcal{H} \rightarrow \mathbb{R}$ to perform parameter updates with gradient based rules such as the given example in Eq.(\ref{eq:grad_desc}). This gives the architecture shown in Fig.{\ref{fig:architectures_b}}, hereby referred to as the vanilla model. We note here that all Ansatz-based HL algorithms such as \cite{wilde2022scalably, gu2022practical, bairey2020learning} can be cast as vanilla models, tuning the $\theta$ according to different Loss functions regardless of whether they focus on experimental friendliness \cite{wilde2022scalably}, or the Black-Box scenario \cite{gu2022practical}. In this langauge, we see these HL algorithms are limiting if there are additional terms not included in $H_A(\theta)$. Furthermore, as we will show in Sec. \ref{sec:results}, using a vanilla HL algorithm causes convergence issues due to the loss landscape of $\theta$, even on a relatively small number of spins.

\subsection{Combining Ansatz with NODEs}
\label{sec:method}
\begin{figure}
\begin{subfigure}{\textwidth}
  \centering
  \begin{tikzpicture}[
    every edge/.style = {draw,->}
  ]
    \node (state) at (0, 0) [draw, minimum width=3cm] {$|\psi_0\rangle$};
  \node (h) at (3, 0) [draw] {$e^{-iH_T t}$};
  \node (times) at (2, -1.8) [draw, circle] {$\times$};
  \node (end) at (2, -3.3) [] {};
  \draw (state) edge[out=270, in=90, looseness=1] (times);
  \draw (h) edge[out=270, in=90, looseness=1] (times);
  \draw (times) edge[] (end);
\end{tikzpicture}
  \caption{}
  \label{fig:architectures_a}
\end{subfigure}

\vspace{10mm}

\begin{subfigure}{\textwidth}
  \centering
  \begin{tikzpicture}[
    every edge/.style = {draw,->}
  ]
  \node (start) at (0, 1.6) [draw, minimum width=3cm] {$|\psi_0\rangle$};
  \node (state) at (0, 0) [draw, minimum width=3cm] {$|\psi\rangle$};
  \node (h) at (3, 0) [draw] {$-iH(\theta)$};
  \node (times) at (2, -1.8) [draw, circle] {$\times$};
  \node (end) at (2, -3.3) [] {};
  \draw (state) edge[out=270, in=90, looseness=1] (times);
  \draw (h) edge[out=270, in=90, looseness=1] (times);
  \draw (times) edge[] (end);
  \draw (start) edge[] (state);
  \node (ode) at (1.1, -0.9) [draw, minimum width=6cm, minimum height=3.5cm, label={[shift={(-2,-3.5)}]ODEint}] {};
\end{tikzpicture}
  \caption{}
  \label{fig:architectures_b}
\end{subfigure}

\vspace{10mm}

\begin{subfigure}{\textwidth}
  \centering
  \begin{tikzpicture}[
    every edge/.style = {draw,->}
  ]
  \node (start) at (0, 1.6) [draw, minimum width=3cm] {$|\psi_0\rangle$};
  \node (state) at (0, 0) [draw, minimum width=3cm] {$|\psi\rangle$};
  \node (h) at (3, 0) [draw] {$-iH(\theta)$};
  \node (times) at (3, -2) [draw, circle] {$\times$};
  \node (mlp) at (-1, -2) [draw, minimum height=1cm, minimum width=3cm] {NN$(\varphi)$};
  \node (add) at (2, -4) [draw, circle] {$+$};
  \node (end) at (2, -5.5) [] {};
  \draw (state) edge[out=270, in=90, looseness=1] (times);
  \draw (h) edge[] (times);
  \draw (state) edge[out=270, in=90, looseness=1] (mlp);
  \draw (times) edge[out=270, in=90, looseness=1] (add);
  \draw (mlp) edge[out=270, in=90, looseness=1] (add);
  \draw (add) edge[] (end);
  \draw (start) edge[] (state);
  \node (ode) at (0.6, -2) [draw, minimum width=7cm, minimum height=5.5cm, label={[shift={(-2.6,-5.5)}]ODEint}] {};
\end{tikzpicture}
  \caption{}
  \label{fig:architectures_c}
\end{subfigure}
\caption{Architectures of the models used to simulate the time evolution of the parameterised Hamiltonian $H(\theta)$. (a) The numerically exact time evolution under the Hamiltonian. (b) The approximate time evolution through integrating the Schrodinger equation, hereby referred to as the vanilla model. (c) The time evolution through integrating the Schrodinger equation with an added corrective term to the right-hand-side.}
\label{fig:architectures}
\end{figure}
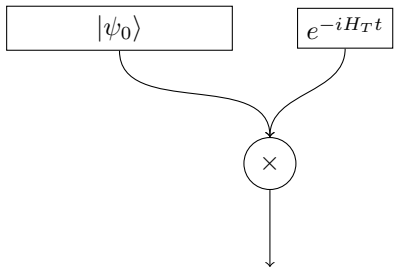
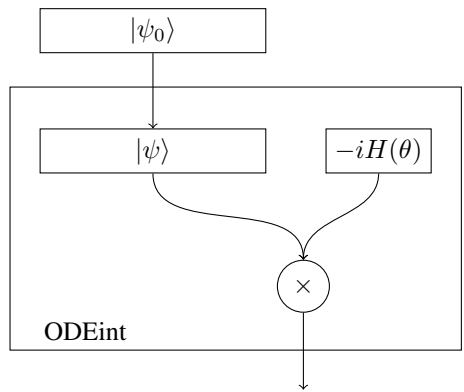
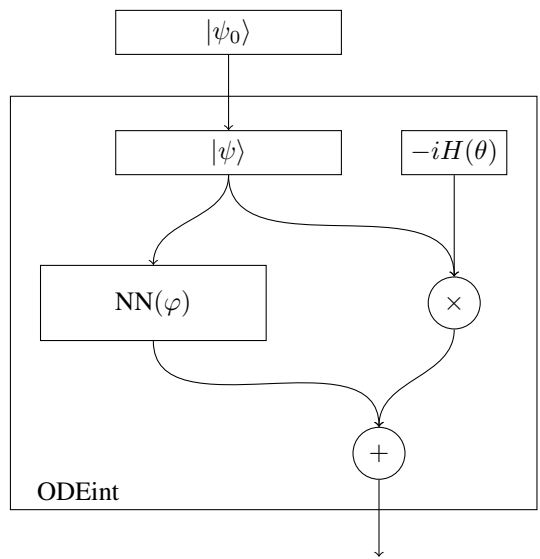

Consider a representation of $H_T$ decomposed as the sum of an Ansatz Hamiltonian, $H_A(\theta)$, and a NODE with a Neural Network (NN) architecture acting as a map $\text{NN}:\mathcal{H} \xrightarrow{\theta} \mathcal{H}$,
\begin{equation}
    H_T = H_A(\theta) + \text{NN}(\varphi),
\end{equation}
with independent, tunable parameters $\theta, \;\varphi$ for the Ansatz and NN respectively. Under this decomposition, the Schrödinger equation reads,
\begin{equation}
    i \ket{\dot{\psi}(t)} = H_{A}(\theta) \ket{\psi(t)} + \text{NN}(\ket{\psi(t)};\varphi),
\end{equation}
which can be expressed as,
\begin{equation}
    \ket{\psi(t + \Delta t)} = 
    \int_t^{t + \Delta t} \frac{d }{dt'} \ket{\psi(t')}dt'
    =
    -i \int_{t}^{t + \Delta t}  H_{A}(\theta) \ket{\psi(t')} dt' -i \int_{t}^{t + \Delta t} \text{NN}(\ket{\psi(t')};\varphi) \; dt'.
\end{equation}
for numerical integration. In integral form, we may represent the Schrödinger equation as a computational graph with a NODE component, as shown in Fig. \ref{fig:architectures_c}. This architecture sits in contrast with the Ansatz-based methods employed by state-of-the-art protocols in \cite{wilde2022scalably, gu2022practical}, shown in Fig.\ref{fig:architectures_b}.

\subsection{Measurements and Loss Function}
\label{sec:measure_and_loss}
In order to provide feedback to the NODE, we must design a function, $\mathcal{L}:\mathcal{H} \times \mathcal{H} \rightarrow \mathbb{R}$, which compares the estimated time-evolved state, $\ket{\psi(t;\theta,\varphi)}$ to the ground truth, $\ket{\psi_T(t)}$, evolved under the action of $H_T$. To make our loss function experimentally friendly, we will assume that access to $\ket{\psi_T(t)}$ can only be achieved by sampling from compatible marginal distributions of $\ket{\psi_T(t)}$, corresponding to Pauli measurements. As such, we will sample from the marginals of the true output state , $\ket{\psi_T(t)}$, giving a dataset, $\mathcal{D}$, comprising binary measurement outcomes for each Pauli basis measurement chosen at random. Using this dataset, we can then compute the log-likelihood that such a dataset can be sampled from the estimated time-evolved state, $\ket{\psi(t;\theta,\varphi)}$. As per \cite{wilde2022scalably}, for $J$ time-stamps, with $M$ repeats per measurement, and $K$ different random Pauli basis, we have $|\mathcal{D}| = MKJ$. Given that we will also vary the input state, this creates a dataset with $|\mathcal{D}| = MKJL$ points, where $L$ is the number of different chosen input states. In order to construct a loss function from this data-set, we note that since our pipeline is numerical, we \textit{do} have full access to the estimator $\ket{\psi(t;\theta,\varphi)}$ of the underlying true output $\ket{\psi_T(t)}$. As such, it is possible to use the Born rule to calculate the probabilities of observing bitstrings, $b$, in the dataset,
\begin{equation}
    p\left( b \; \big| \ket{\psi_{\theta}(t)} \right) =  |\braket{b |\psi_{\theta}(t)}|^2.
\end{equation}
From here, we can use the probabilities to construct our loss function as the average negative log-likelihood of these outcomes,
\begin{equation}
    L(\ket{\psi_{\theta}}; \mathcal{D})= -\sum_{b \in \mathcal{D'}} \log |\braket{{b}|{\psi_\theta}}|^2,
\end{equation}
where $\mathcal{D'} \subset \mathcal{D}$ is a subset of the dataset corresponding to bitstrings acquired from evolving the correct input state, and measuring at the same timestamp as the estimator. 

This enables us to create an update rule for the ODE via back propagation of the loss function to the trainable parameters, $\theta$ and $\phi$ of the Ansatz Hamiltonian and NN, respectively. \\

Depending on the NN architecture, the NN component of the architecture in Fig. \ref{fig:architectures_c} can be much more expressive that $H_A(\theta)$, meaning it is powerful enough to overfit the training data and completely ignore the contribution of $H_A(\theta)$. For example, a Multi-Layer Perceptron (MLP), which has fully connected layers, is able to correlate any element of an input wavefunction with any other. Physically, this would correspond to an action which couples every spin to every other spin in our many-body system. This is of concern since it is beneficial for our representation to be \textit{intepretable} in the white-box scenario from Fig.\ref{fig:white_box_black_box} - which is the case when as much of the 
output of Fig.\ref{fig:architectures_c} as has large a contribution as possible from  the analytic component, $H_{A}(\theta)$, rather than the NN. Hence, we should seek to make the architecture be as dependent on the ansatz component as possible. We may address this problem with heavy regularisation of the NN \cite{girosi1995regularization}, and with a curriculum learning strategy \cite{bengio2009curriculum, soviany2022curriculum} discussed below. In the black-box scenario of Fig.\ref{fig:white_box_black_box}, we can ourselves make an ansatz and use curriculum learning in order to estimate the extent to which our ansatz is correct. Further details can be found in Sec. \ref{sec:curriculum_learning}.

\subsection{Curriculum learning}
\label{sec:curriculum_learning}
As stated above, the expressive power of the NN means its contribution to the output state could dominate over the Ansatz Hamiltonian. The reason for this is that the NN receives as input a vectorised state and is fully connected. Thus, through the hidden, fully connected layers, the NN is able to enact transformations that couple state coefficients that would otherwise be uncoupled with just the Ansatz Hamiltonian alone\footnote{Assuming that the Ansatz is an element of a subgroup of $\mathcal{P}_N$, otherwise it has an exponentially large number of parameters rendering it untrainable for large system sizes.}, as well as being able to couple the same coefficients of the input state, $\ket{\psi_T(t = 0)}$, as the Ansatz Hamiltonian. Consequently, the NN is in principle able to dominate over the Ansatz Hamiltonian, causing complications in the white-box scenario in Fig.\ref{fig:white_box_black_box}. However, if we seek an analytic answer to the HL problem (as opposed to simply finding a neural representation of it, solving the black-box scenario with no interpretation), we wish for as much of the output as possible to come from the Ansatz. To enable this, we use the principle of curriculum learning. In short, curriculum learning is a training strategy that gradually increases in complexity as training progresses. To that end, we use the following curriculum
\subsection*{Curriculum 1}
\begin{itemize}

\item [i.] Warm-up phase: the Hamiltonian parameters, $\theta$, and the NN parameters, $\phi$, are trained simultaneously.

\item [ii.] First phase: the NN is \textit{switched off}\footnote{Note that this is different from freezing the NN parameters, $\phi$.} for both the forward and backward passes. The Hamiltonian parameters, $\theta$, are trained.

\item [iii.] Second phase: the Hamiltonian parameters, $\theta$ are frozen and the NN's parameters, $\phi$, are trained.

\end{itemize}

Our reasoning for the warm-up phase comes from the fact that works such as \cite{wilde2022scalably, gu2022practical, franca2022efficient} only converge to the correct ground truth for specific parameter initialisations. This sensitivity to initial parameters is due to a barren loss landscape, local minima, or insensitivity of the loss function to variation of the parameters in these initialisation regions. See Sec. \ref{sec:results} and Fig.\ref{fig:landscape} for further discussion. In practise, we found that having a warm-up phase avoids this problem entirely. Further discussion on the curriculum and initialisation can be found in Sec. \ref{sec:robustness} where we directly compare the robustness to the results in  \cite{wilde2022scalably}. Finally, we note that for the case where Curriculum 1 fails to be robust, we can further break the HL problem down into smaller parts. Since our proof of concept up to $N = 8$ spins converges with Curriculum 1 for our test set detailed below, we reserve a brief discussion on this extended curriculum to Appendix \ref{app:curriculum_2}. \\

% \textcolor{red}{Can we cut this section? We never used it in our results section - and did not test sufficiently hard Hamiltonians ($N$ was limiting here probably) to make this component necessary). We could also put it in the discussion section to discuss how to maintain high accuracy as we scale. Lastly, if we cut this, in the discussion section we should indicate how the parameters converging to the ground truth implies the neural component outputs $0$, and that the percentage that the architecture relies on the neural network component tells us how good/bad the guess was. From Marcin's comment, if the Ansatz is the clifford group, then it also gives us some measure of non-simulability!}

\section{Results}
\label{sec:results}
In this section, we benchmark the performance our algorithm on 1D spin-$1/2$ chains for, $N \in \{3,\ldots,8\}$ spins. We begin by explaining the datasets used in training, detailing the Hamiltonians tested, and neural network architecture. We choose a wide variety of Hamiltonians which include higher order polynomials of Pauli operators, and non-ergodic systems \cite{ho2018ergodicity, prosen2002general}. We also choose Hamiltonians from state-of-the-art works such as \cite{wilde2022scalably, gu2022practical} in order to benchmark our work against current methods. We then probe the robustness of our algorithm by evaluating its success rates on our test Hamiltonians, and evaluate its performance outside of training times using our new benchmark defined in Sec. \ref{sec:benchmark}. Finally, we analyse our algorithm's limitations and experimental friendliness.

\subsection{Datasets and Ground Truth}\label{sec:dataset}
To find some reasonable test Hamiltonians, we consider the work of \cite{wilde2022scalably}, where the authors use a 1D Heisenberg spin chain, with homogeneous coupling coefficients and a local transverse field,
\begin{equation}
    H = \sum_{i=1}^{N-1}\left[J^xX_iX_{i+1}+J^yY_iY_{i+1}+J^zZ_iZ_{i+1}\right]+\sum_{i=1}^Nh_iX_i
    \label{eq:isotropic_heis_ham},
\end{equation}
where the number of spins, $n$, ranges between 6 and 100. On the other hand, the authors of \cite{gu2022practical} use a transverse field Ising model with 9 spins as their example.
We test our method on the 1D Heisenberg spin chain from \cite{wilde2022scalably}, as well as the following modifications to the Heisenberg Hamiltonian to probe the capabilities of our method on more complex Hamiltonians:
\begin{enumerate}
\item The Anisotropic Heisenberg Hamiltonian:
\begin{equation}
    H_1 = \sum_{i=1}^{N-1}\left[J_i^xX_iX_{i+1}+J_i^yY_iY_{i+1}+J_i^zZ_iZ_{i+1}\right]+\sum_{i=1}^Nh_iX_i
    \label{eq:anisotropic_heis_ham}
\end{equation}

\item The next-nearest-neighbour Heisenberg Hamiltonian:

\begin{align}
H_2&=\sum_{i=1}^N\left[h^x_iX_i+h^y_iY_i+h^z_iZ_i\right]+\sum_{i=1}^{N-1}\left[J_i^xX_iX_{i+1}+J_i^yY_iY_{i+1}+J_i^zZ_iZ_{i+1}\right]\nonumber\\
&\qquad\qquad+\sum_{i=1}^{N-2}\left[K_i^xX_iX_{i+2}+K_i^yY_iY_{i+2}+K_i^zZ_iZ_{i+2}\right]
\label{eq:nnn_ham}
\end{align}

\item The third-order Heisenberg Hamiltonian:

\begin{align}
H_3&=\sum_{i=1}^N\left[h^x_iX_i+h^y_iY_i+h^z_iZ_i\right]+\sum_{i=1}^{N-1}\left[J_i^xX_iX_{i+1}+J_i^yY_iY_{i+1}+J_i^zZ_iZ_{i+1}\right] \nonumber \\
&\qquad\qquad+\sum_{i=1}^{N-2}\left[K_i^xX_iX_{i+1}X_{i+2}+K_i^yY_iY_{i+1}Y_{i+2}+K_i^zZ_iZ_{i+1}Z_{i+2}\right]
\label{eq:third_order_ham}
\end{align}

\item The PXP Hamiltonian \cite{serbyn2021quantum, liu2024quantum}, which is one of the simplest examples of a non-ergodic Hamiltonian:
\begin{align}
    H_4=\sum_{i=2}^{N-1}J_iP_{i-1}X_iP_{i+1}
    \label{eq:pxp_ham}
\end{align}
where $P_k=\ket{0}\bra{0}_k$ is the projector onto the $\ket{0}$ state for the $i$-th spin.

\item The dense nearest-neighbour Hamiltonian:
\begin{equation}
    H_5=\sum_{i=1}^N\left[h^x_iX_i+h^y_iY_i+h^z_iZ_i\right]+\sum_{i=1}^{N-1}\sum_{A,B}\left[J_i^{AB}A_iB_{i+1}\right]
    \label{eq:dense_nn_ham}
\end{equation}
$$$$
where in the second sum, $A$ and $B$ are each taken over the Pauli matrices $\{X,Y,Z\}$.

\end{enumerate}

We generate our ground-truth dataset using an exact state-vector simulation of the system evolving under the true Hamiltonian, whose computational graph is shown in Fig.\ref{fig:architectures_a}. To simplify training data collection from hardware, we can use short interaction times, ensuring the system's coherence isn't needed for long periods.

For all system sizes $N\in\{3,4,5,6,7,8\}$, we use $L=5$ randomly chosen initial computational basis states, $J=5$ timestamps corresponding to $t\in\{0.2, 0.4, 0.6, 0.8, 1.0\}$ (in units of $\hbar$), $K=200$ Pauli bases chosen at random from the uniform distribution of Pauli strings, and $M=100$ shots per each combination of state, timestamp, and Pauli measurement. This means that for fewer than $4$ bodies, the Pauli measurements \textit{can be} informationally complete, but not necessarily since the Pauli-basis measurements are sampled at random. For all datasets higher than $4$, the number of shots limits the measurement set to \textit{always} be informationally incomplete.

For each of our Hamiltonians, we sample the true parameters, $c_j$ uniformly from the interval $[-1,1]$ to test the robustness over 50 different samples. This interval assumes that all parameters are of the same order, however our method can also apply when true parameters vary over more orders of magnitudes. We note here that this means our proof of concept does not consider Hamiltonians with critical points and phase transitions. However, since $N = 8$ bodies in our study, effects such as power-law decays, spin-spin correlations, gapped and gapless phases and different excitation spectra do not dominate dynamics for this relatively small number of bodies \cite{tasaki2020physics, thouless2014quantum}.

In order to estimate the generalisation capabilities of our algorithm, we define the Infidelity Loss of an output quantum state,
\begin{equation}
    L\big(\ket{\psi_T(t)}, \ket{\psi(t;\theta)}\big) = 1 - \mathcal{F}
    \big(\ket{\psi_T(t)}, \ket{\psi(t;\theta)}\big),
\end{equation}
where $\mathcal{F}$ is any fidelity function, in our case we use the overlap, $\mathcal{F} = |\braket{\psi_1|\psi_2}|^2$.

We can use this loss on testing datasets by allowing our system to evolve past its training times. For this test set, we use the same ground-truth parameters across all experiments for comparison purposes, tracking 5 different basis states. \\

We now proceed to understand the performance of our algorithm in terms of three aspects: robustness, accuracy and extrapolation behaviour, and experimental friendliness.
\subsection{Robustness}
\label{sec:robustness}
To test the robustness of our algorithm, we can initialise each of models described on Eqs.(\ref{eq:isotropic_heis_ham}-\ref{eq:dense_nn_ham}) for different ground truth initialisations. We may then evaluate the robustness by estimating how often our algorithm converges for different ground truth parameters. To that end, let $\theta$ be the set of ground-truth parameters for Hamiltonians in our test set Eqs.(\ref{eq:isotropic_heis_ham}-\ref{eq:dense_nn_ham}). 
For example, the Anisotropic Heisenberg Hamiltonian in Eq.(\ref{eq:anisotropic_heis_ham}) has $\theta = \{J_i^x, J_i^y, J_i^z, h_i^z\},\;\; i \in \{1,\ldots,N\}$, 
whilst $\theta = \{h_i^x,h_i^y,h_i^z,
J_i^x,J_i^y,J_i^z,K_i^x,K_i^y,K_i^z\},\;\;i\in
\{1,\ldots N\}$
for the next-nearest-neighbour Heisenberg Hamiltonian of Eq.(\ref{eq:nnn_ham}).
Furthermore, let $\tilde{\theta}$ be the set of estimators for the ground-truth parameters for Hamiltonians in our test set, Eqs.(\ref{eq:anisotropic_heis_ham}-\ref{eq:dense_nn_ham}). We may then define the relative absolute error between the true Hamiltonian parameters, $\theta$, and the predicted parameters, $\tilde{\theta}$, as
\begin{equation}
\epsilon(\tilde{\theta})=\frac{\|\theta-\tilde{\theta}\|}{\|\theta\|},
\label{eq:relative_error}
\end{equation}
where $\|\cdot\|$ is the L1 norm. 

We say that the learning algorithm is successful if the relative absolute error is lower than some threshold, which we set to approximately 0.1. This choice of threshold is inconsequential since in all experiments the separation of the relative errors between successful trials and unsuccessful trials is large. We chose 0.1 because it is approximately the noise rate of state-of-the-art hardware's two-qubit gate infidelity \cite{decross2024computational}.

\begin{table}
\centering
\begin{tabular}{|c|c|c|c|}\hline
    Hamiltonian & $N$ & Vanilla & Neural ODE \\ \hhline{|=|=|=|=|}
    & 3 & 66\% & 100\% \\ \cline{2-4}
    & 4 & 86\% & 100\% \\ \cline{2-4}
    Heisenberg (Eq.\ref{eq:isotropic_heis_ham}) & 5 & 94\% & 100\% \\ \cline{2-4}
    & 6 & 64\% & 100\% \\ \cline{2-4}
    & 7 & 52\% & 100\% \\ \cline{2-4}
    & 8 & 52\% & 100\% \\ \hline
    & 3 & 28\% & 96\% \\ \cline{2-4}
    & 4 & 46\% & 96\% \\ \cline{2-4}
    Anisotropic & 5 & 34\% & 94\% \\ \cline{2-4}
    Heisenberg (Eq.\ref{eq:anisotropic_heis_ham})& 6 & 32\% & 96\% \\ \cline{2-4}
    & 7 & 18\% & 96\% \\ \cline{2-4}
    & 8 & 16\% & 96\% \\ \hline
    & 3 & 12\% & 100\% \\ \cline{2-4}
    & 4 & 12\% & 100\% \\ \cline{2-4}
    PXP (Eq.\ref{eq:pxp_ham}) & 5 & 6\% & 100\% \\ \cline{2-4}
    & 6 & 2\% & 100\% \\ \cline{2-4}
    & 7 & 0\% & 98\% \\ \cline{2-4}
    & 8 & 0\% & 94\% \\ \hline
% \end{tabular}\qquad
% \begin{tabular}{|c|c|c|c|}\hline
    % Hamiltonian & $N$ & Vanilla & Neural ODE \\ \hhline{|=|=|=|=|}
    & 3 & 94\% & 100\% \\ \cline{2-4}
    & 4 & 98\% & 100\% \\ \cline{2-4}
    Dense NN (Eq.\ref{eq:dense_nn_ham}) & 5 & 96\% & 98\% \\ \cline{2-4}
    & 6 & 98\% & 98\% \\ \cline{2-4}
    & 7 &  88\% & 98\%\\ \cline{2-4}
    & 8 &  98\% &  98\% \\ \hline
    & 3 & 18\% & 98\%\\ \cline{2-4}
    & 4 & 56\% & 100\%\\ \cline{2-4}
    Heisenberg NNN (Eq.\ref{eq:nnn_ham}) & 5 &  48\% & 92\% \\ \cline{2-4}
    & 6 &  50\% & 86\% \\ \cline{2-4}
    & 7 & 46\% &  96\% \\ \cline{2-4}
    & 8 & 56\% & 94\% \\ \hline
    & 3 & 54\% & 100\% \\ \cline{2-4}
    & 4 & 68\% &  100\% \\ \cline{2-4}
    3rd order & 5 & 70\% & 100\% \\ \cline{2-4}
    Heisenberg (Eq.\ref{eq:third_order_ham}) & 6 & 68\% &  86\% \\ \cline{2-4}
    & 7 & 78\% & 96\%  \\ \cline{2-4}
    & 8 & 64\% & 94\% \\ \hline
\end{tabular}
\medskip
\caption{Success rates of the vanilla model in \cite{wilde2022scalably} and the proposed neural ODE model for the 6 Hamiltonians indicated, calculated over 50 runs with random parameter initialization.}
\label{tab:success_rate}
\end{table}

\begin{figure}
    \centering
\includegraphics[width=\textwidth]{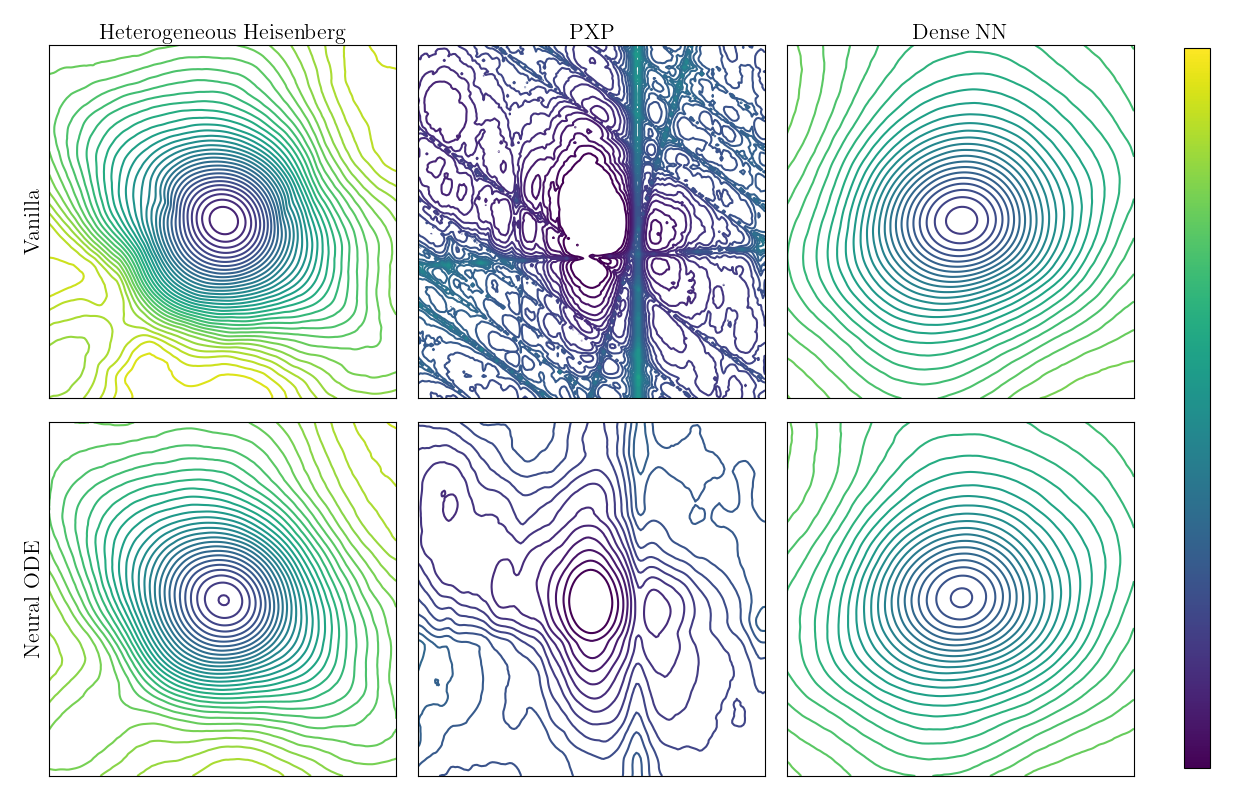}
    \caption{The loss landscapes of for two parameters in the anisotropic Heisenberg Hamiltonian, the PXP Hamiltonian, and the dense nearest-neighbour Hamiltonian respectively from left to right. The center of each plot is the global minimum, and the axes are chosen as two random orthogonal directions in the parameter space. The top row contains the landscapes for the vanilla model architecture from \cite{wilde2022scalably}, shown in Fig.\ref{fig:architectures_b}. The bottom row contains landscapes for the Neural ODE model from Fig.\ref{fig:architectures_c}, over the same Hamiltonians, with the same ground-truth parameters. We see clearly in the centre two plots that the Neural ODE model can make significant improvements to the loss landscape in terms of the number of local minima, as well as having a smoothing effect to the ruggedness of the PXP Hamiltonian's landscape. This effect is visible in Table \ref{tab:success_rate}, where the convergence rate of the Neural ODE model greatly surpasses the vanilla model. Notice that for the left panels (Heterogeneous Heisenberg), the Neural ODE model has changed the loss landscape in the lower-left corner to have fewer saddle points and a more even loss landscape. This makes it more likely to approach the global minimum when optimising as there are more directions from which one can arrive at the local minimum. However, since the change is relatively small, the Neural ODE causes only a small boost in the success rate, as shown in Table \ref{tab:success_rate}. Finally, notice that for the right panels (Dense NN), the Neural ODE makes almost no change to the loss landscape near the global minimum. We see this reflected in the almost equivalent success rates of the vanilla model vs the Neural ODE model in Table \ref{tab:success_rate}.
    }
    \label{fig:landscape}
\end{figure}

The success rates of the vanilla model and the neural ODE model on the 6 Hamiltonians above are shown in table (\ref{tab:success_rate}). We see that the vanilla model has a significant probability of failure across our test set when compared to the neural augmentation's performance. Considering the loss landscape of these two models in Fig.\ref{fig:landscape}, we infer that this this is due to their difference in sensitivity to parameter initialization. In the vanilla model from Fig.\ref{fig:architectures_a} (with only the Ansatz parameters), the loss landscape is more barren and ridden with local minima, thus making it difficult for effective gradient-based parameter tuning as shown in Fig.\ref{fig:landscape}. However, the neural ODE model from Fig.\ref{fig:architectures_c} has success rates close to 100\% for all of the Hamiltonians evaluated up to $N = 8$ bodies. Considering Fig.\ref{fig:landscape}, which shows how the loss landscape changes under the neural augmentation of Fig. \ref{fig:architectures_c}, this neural augmentation improves the loss landscape to one with fewer local minima and barren plateaus. Thus we see that adding the neural augmentation helps to remove the sensitivity of the architecture to initialization of its parameters. In a smoother loss landscape, there are more initialisations of the Ansatz parameters which will
converge towards the global optimum. 

This is further reinforced by the fact that the vanilla model is much more robust on Hamiltonians such as the dense nearest-neighbour landscape of Eq.(\ref{eq:dense_nn_ham}), whose loss landscape appears smooth with all gradients pointing almost directly to the global optimum. In this instance, the Neural ODE model makes little change to the loss landscape near the global minimum, meaning the vanilla model and Neural ODE model have similar performance for this Hamiltonian. Whereas for the PXP Hamiltonian of Eq.(\ref{eq:pxp_ham}), the ruggedness of the loss landscape in Fig.\ref{fig:landscape}
explains the almost vanishing success rate of the vanilla model in learning the PXP Hamiltonian.

\subsection{Accuracy and Extrapolation Behaviour via New Benchmark}
\label{sec:benchmark}

\begin{figure}
    \centering
    \includegraphics[width=\textwidth]{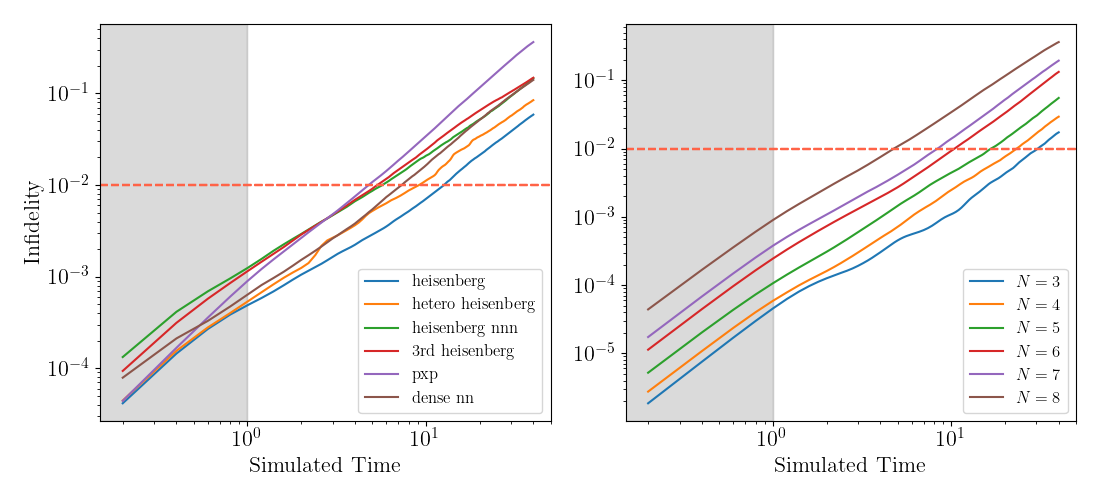}
    \caption{Infidelities between the simulated time-evolved state of the neural ODE and the true time-evolved state. The boundary of the shaded region on the left represents the training time (i.e. feeding back at this boundary point only), and to the right of it are testing times. Here, the red dashed line marks the 1\% error boundary, past which the model has more than a 1\% error. (Left) The fidelities as a function of the simulation time for the 6 Hamiltonians indicated, where here we fix the number of spins to $N=8$. (Right) The fidelities as a function of the simulation time for different system sizes, where here we use the PXP Hamiltonian as this was heuristically found to be the most difficult as per Table \ref{tab:success_rate}.
    }
    \label{fig:ode_fidelity_size}
\end{figure}

We can also evaluate the performance of our algorithm by testing how it generalises to times unseen in its training. Here, we focus exclusively on the neural ODE model, as the vanilla model lacks the robustness needed to reliably learn more complex Hamiltonians in the test set Eqs.(\ref{eq:isotropic_heis_ham} - \ref{eq:dense_nn_ham}). For this purpose, we generate random initial computational basis states not seen during training, and perform a simulation up to $T=20$ of the learned neural ODE model ($20\times$ the time used in training). We do this for each of the different Hamiltonians and system sizes, keeping the ground truth parameters fixed. We note that this is different from the test set used to establish robustness, where the ground truth parameters were varied in order to ensure convergence over a range of different true parameters. The results of this testing are shown in Fig. (\ref{fig:ode_fidelity_size}). 

We see that the neural ODE model is able to generalize to longer simulation times than seen in training. For up to $N = 8$ spins and over the Hamiltonians described in Sec. \ref{sec:dataset}, Fig.\ref{fig:ode_fidelity_size} shows a fidelity of approximately $99\%$ for ten times the training times seen. More interestingly, we observe from Fig. (\ref{fig:ode_fidelity_size}) that the infidelity as a function of the simulation time can be fitted to a power law:
\begin{equation}
1-F\big(\ket{\psi(t)},\ket{\psi'(t)}\big)=At^b,\label{eq:fid_power_law}
\end{equation}
where $\ket{\psi(t)}$ and $\ket{\psi'(t)}$ are the ground-truth and predicted states at time $t$, respectively, and $A$ and $b$ are positive constants whose values depend on the particular Hamiltonian and system size. This power law will start to fail at the point when the predicted state diverges from the true state. However, Fig.\ref{fig:ode_fidelity_size} shows that for our model, the power law holds for sufficiently small $t$, at least up to twenty times the evolution time seen during training.

The coefficients $A$ and $b$ of the power law can be determined empirically through linear regression. For the numerical experiments in Fig.(\ref{fig:ode_fidelity_size}), the power law coefficients are shown in Table (\ref{tab:power_law}). The power law in time past the training region in Fig.\ref{fig:ode_fidelity_size} shows our algorithm has a global error (measured in terms of the Infidelity Loss from Eq.(\ref{eq:fid_power_law})) which empirically scales polynomially with an exponent $1 \leq b \leq 2$ for all the Hamiltonians tested. This indicates that our algorithm can generalise with at most a quadratic amount of error in time for the tested examples, implying that the neural augmentation indeed does form a representation of the unitary dynamics, rather than overfitting to the training data. Furthermore, the accuracy of the exponent, in terms of the error values of Table (\ref{tab:power_law}), shows that our method can generalise beyond the training time with a predictable amount of uncertainty in its estimates for up to twenty times the training data point. \\

\subsubsection*{Power Laws can Benchmark Hamiltonian Learning Algorithms}

Despite the variety of approaches to perform HL in diverse settings, there is not yet a systematic, quantitative means of comparison for HL algorithms that is agnostic to their inner workings. Whilst some metrics of Hamiltonian complexity such as the so-called V-score \cite{wu2023variational} have been defined, the complexity measure is as estimate of simulation hardness. This makes metrics like the V-score appropriate in the black-box scenario of Fig.\ref{fig:white_box_black_box}, as the main use-case of black-box Hamiltonian Learning is to construct a simulation of the ground truth. In contrast, the V-score does not take into account the difficulty in tuning a known structure, as per the white-box scenario in Fig.\ref{fig:white_box_black_box}. Nor does it consider the sensitivity of successful convergence to a HL algorithm's initial conditions. This is because we are trying to \textit{infer} a given system's unknown Hamiltonian from data, rather than attempt to simulate a known Hamiltonian. Therefore, users and researchers of HL currently face difficulties in deciding which algorithm best fits their needs. \\

We believe that the scaling behaviour of Eq.(\ref{eq:fid_power_law}) provides a reasonable benchmark for the ability of a generic HL algorithm to generalise outside training data. This is because the exponent of the power law shows two important aspects of generalisation. First, a lower value for the exponent implies that a HL algorithm can maintain accuracy for longer periods in time in a dataset unseen by training. Second, an exponent with low error means we can reliably understand the amount of uncertainty in the estimator of the output. With a higher error in the exponent of Eq.(\ref{eq:fid_power_law}), we see a breakdown in the generalisation capabilities of a HL algorithm; being unable to rely on the uncertainty in the scaling behaviour means the algorithm has high variability in its ability to converge, and is not guaranteed to remain accurate outside its training data. There are several advantages of using power-laws to benchmark HL algorithms. First, Fig.\ref{fig:ode_fidelity_size} shows the scaling law was observed to have similar coefficients over a diverse set of 1D spin-$1/2$ chains. As such, we assert that the $b$ coefficient and its error are indicative of our HL algorithms's performance beyond our examples from Eqs.(\ref{eq:isotropic_heis_ham} - \ref{eq:dense_nn_ham}), and can help future users gauge how well our HL algorithm could perform on different examples not covered in this manuscript. Second, seeing a similar range of $b$ values independently of the number of bodies, indicates that our protocol is scalable should we be able to tame the exponential scaling of the input state vector (See Sec. \ref{sec:conclusion} for further discussion). These factors indicate that the range of values of $b$ is a property of the algorithm, rather than the ground-truth examples tested. This is important because it distinguishes this benchmark from the so-called V-score \cite{wu2023variational}, which is dependent on the Hamiltonian to be learned by construction. Finally, we note that we can calculate the $b$ coefficient for \textit{any} HL algorithm. Regardless of its intricacies, a HL algorithm should return an estimate for the true Hamiltonian. With this estimate, we can choose some test states unseen during training and evolve them in time past the training region. This is possible for any HL algorithm because the only three objects needed are the initial states, the true Hamiltonian, and the estimated Hamiltonian. All of these are available for both scenarios shown in Fig.\ref{fig:white_box_black_box}.

\begin{table}
\centering
\begin{tabular}{|c|c|c|}\hline
    Hamiltonian & $A$ & $b$ \\ \hhline{|=|=|=|}
    Heisenberg & $3.56\times 10^{-4}\pm 7.66\times 10^{-6}$ & $1.364\pm 7.42\pm 10^{-3}$ \\ \hline
    Anisotropic Heisenberg & $5.15\times 10^{-4}\pm 5.56\times 10^{-6}$ & $1.384\pm 3.74\times 10^{-3}$ \\ \hline
    PXP & $7.56\times 10^{-4}\pm 6.74\times 10^{-6}$ & $1.678\pm 3.10\times 10^{-3}$ \\ \hline
    Dense NN & $5.49\times 10^{-4}\pm 8.75\times 10^{-6}$ & $1.504\pm 5.51\times 10^{-3}$ \\ \hline
    Heisenberg NNN & $1.10\times 10^{-3}\pm 1.30\times 10^{-5}$ & $1.299\pm 4.09\times 10^{-3}$ \\ \hline
    3rd order Heisenberg & $1.09\times 10^{-3}\pm 1.07\times 10^{-5}$ & $1.347\pm 3.38\times 10^{-3}$ \\ \hline
\end{tabular}

\medskip

\begin{tabular}{|c|c|c|}\hline
    System size & $A$ & $b$ \\ \hhline{|=|=|=|}
    3 & $2.93\times 10^{-5}\pm 9.31\times 10^{-7}$ & $1.686\pm 1.09\times 10^{-2}$ \\ \hline
    4 & $4.72\times 10^{-5}\pm 6.10\times 10^{-7}$ & $1.720\pm 4.47\times 10^{-3}$ \\ \hline
    5 & $9.02\times 10^{-5}\pm 1.38\times 10^{-6}$ & $1.705\pm 5.28\times 10^{-3}$ \\ \hline
    6  & $1.80\times 10^{-4}\pm 3.81\times 10^{-6}$ & $1.758\pm 7.30\times 10^{-3}$ \\ \hline
    7 & $2.73\times 10^{-3}\pm 5.52\times 10^{-6}$ & $1.759\pm 6.97\times 10^{-3}$ \\ \hline
    8 & $7.56\times 10^{-3}\pm 6.74\times 10^{-6}$ & $1.678\pm 3.10\times 10^{-3}$ \\ \hline
\end{tabular}
\medskip
\caption{Empirically-determined coefficients of the power law in equation (\ref{eq:fid_power_law}) with their standard deviations. (Top) The coefficients for the different Hamiltonians with the system size fixed to $N=8$. (Bottom) The coefficients for different system sizes with the Hamiltonian fixed as the PXP model.}
\label{tab:power_law}
\end{table}

\subsection{Experimental Friendliness}

We can evaluate the experimental friendliness of our algorithm by understanding how much training data is required to reach a given relative error described in Eq.(\ref{eq:relative_error}). This is important to ensure that our framework is experimentally friendly, since if the number of Pauli measurements required to effectively learn the Hamiltonian is too large, then the ground-truth dataset would be difficult to generate and the framework would not be useful in practice. 

\begin{figure}[h]
  \centering
  \includegraphics[width=0.8\textwidth]{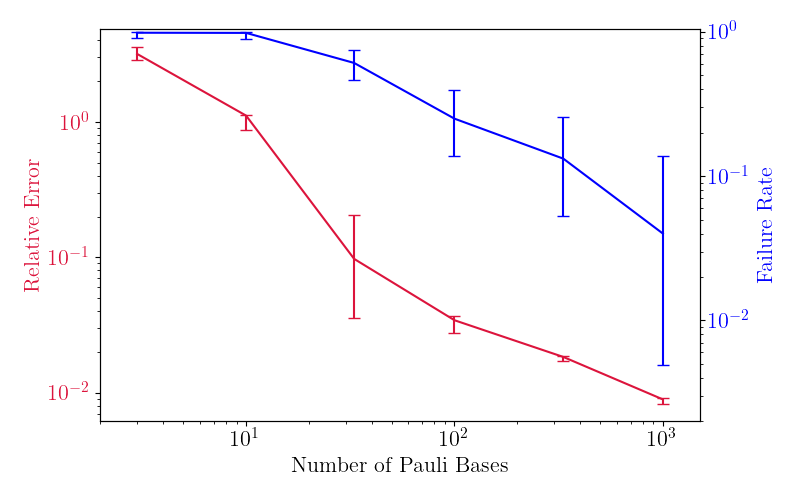}
\caption{The impact of the number of Pauli measurements on the robustness and the relative error of the learning algorithm using the neural ODE model. Here we use the PXP Hamiltonian on a system of 8 bodies to give a worst-case scaling as Table \ref{tab:success_rate} shows this was the most difficult Hamiltonian considered in our examples from Eq.(\ref{eq:isotropic_heis_ham} - \ref{eq:dense_nn_ham}). }
\label{fig:pauli_bases}
\end{figure}

Since our set-up in Fig.\ref{fig:white_box_black_box} does not allow any optimal control \cite{dutkiewicz2023advantage}, we expect the scaling behaviour for the amount of data needed to reach a given relative error to be governed by the Standard Quantum Limit. To this end, we vary the number of Pauli measurements $K\in\{3,10,30,100,300,1000\}$ and keep the rest of the dataset parameters the same as Sec. \ref{sec:dataset}. We evaluate our neural ODE model on the PXP Hamiltonian given in Eq.(\ref{eq:pxp_ham}) on a system with $N = 8$ spins, since this Hamiltonian was empirically found to be the most difficult according to Table (\ref{tab:success_rate}), giving a worst-case-scenario scaling behaviour. The results can be seen in Fig.\ref{fig:pauli_bases}, which evaluates the learning algorithm by observing how the robustness and the relative error scale with the number of Pauli measurements used.

Owing to the results of \cite{dutkiewicz2023advantage}, we see an approximate power law scaling past a certain minimum threshold number of measurements. In this instance for the PXP Hamiltonian of Eq.(\ref{eq:pxp_ham}), we see qualitatively that this is around 30 Pauli bases. This means that we can predictably improve the robustness and the accuracy of the learning algorithm by increasing the number of Pauli measurements. More importantly, the fact that we do not see a power law for a small number (<30 in Fig.\ref{fig:pauli_bases}) of Pauli measurements indicates that there is some minimum number of Pauli measurements in order for the learning algorithm to be reliable. We also note here that changing the measurement process to one with limited or total quantum control could allow us to surpass the standard quantum limit \cite{giovannetti2004quantum, garcia2021learning, dutkiewicz2023advantage}.

\section{Conclusion and Outlook}
\label{sec:conclusion}
In this contribution, we introduced the foundational concepts for how to use Neural ODEs to solve the HL problem on quantum many-body spin systems, as well as preliminary results that highlight its reliability and expressibility on a number of previously unsolved HL problems in 1D spin-$1/2$ chains. In terms of performance, we believe Neural Differential Equations open new avenues for solving the HL problem, which will allow researchers to better characterise a more diverse set of many-body quantum systems. Due to its innate interpretability, we focused on the white-box scenario. However, as discussed in Sec. \ref{sec:curriculum_learning}, our architecture is quite capable of forming a black-box representation when the NODE component dominates over a Hamiltonian ansatz. From the perspective of quantum simulation, our work can hence be used in both the black-box and white-box scenarios. For error mitigation, our work can be used to learn the free Hamiltonian of quantum hardware, allowing users to systematically diagnose the most likely Hamiltonian in their system, and correct for any unexpected interactions. This also applies to those wishing to perform device certification. By using our HL framework on controlled quantum operations, we can verify that they are working correctly by inferring the Hamiltonian maximising the log-likelihood of observed data. In all four of these uses, we believe our new benchmark offers the most informative assessment of the quality of HL algorithms. This is because this benchmark can be considered as a property of the algorithm rather than the Hamiltonian to be learned. To perform the aforementioned tasks, we need to know how well a neural architecture works outside its training times on unseen basis states. The coefficient, $b$, from Eq.(\ref{eq:fid_power_law}) captures these aspects in a fine grained way, irrespective of the V-Score \cite{wu2023variational} of the Hamiltonian to be learned. 

However, we note here that our work has three main challenges to facilitate broader use of HL algorithms. First, our method works in the state-vector formalism of unitary quantum dynamics. This means that the Neural ODE's input size scales exponentially with the number of bodies. Second, the loss function relies on being able to accurately estimate the log-likelihood of a small number of Pauli strings, potentially inducing biases to the loss function if the number of measurement bases is not sufficient, or in the case of noisy measurements. An interesting research direction comes from the idea of tracking expectation values in time, for example via techniques like classical shadows \cite{huang2020predicting}, to address this. Finally, being in the NISQ era of quantum technology, our algorithm should be able to perform reliably in noisy dynamical settings, encapsulated by open-system quantum dynamics, rather than the unitary case. As the first paper in a planned series, the next works aim to systematically investigate the extent to which Neural ODEs can be used to solve the Hamiltonian Learning problem in these settings, addressing the above limitations. \\

\section*{Acknowledgments}
We would like to extend our thanks to Frederik Wilde, Marcin Plodzien, Mark Thomas, Jose Ramon Martinez for their insightful discussions and feedback on the manuscript.
This project was supported by Government of Spain (Severo Ochoa CEX2019-000910-S, Quantum in Spain, FUNQIP and European Union NextGenerationEU PRTR-C17.I1), the European Union (PASQuanS2.1, 101113690 and Quantera Veriqtas), Fundació Cellex, Fundació Mir-Puig, Generalitat de Catalunya (CERCA program), the ERC AdG CERQUTE and the AXA Chair in Quantum Information Science.

%Bibliography
\bibliographystyle{unsrt}  
\bibliography{references}

\newpage
\section{Appendix: Extended curriculum for Hamiltonians with higher complexity}
\label{app:curriculum_2}
% \toni{I WOULD MOVE SUCH SHORT APPENDIX TO THE MAIN TEXT, IF POSSIBLE AND NOT TOO DISTRACTING.}
As mentioned in Sec. \ref{sec:background}, a general Hamiltonian $H_T \in \mathcal{P}_N$ has $4^{N-1}$ coefficients which quickly becomes infeasible as the number of spins grows. To render this problem tractable, we seek to work in a setting where the ground truth is guaranteed to have $\mathcal{O}(M) < N$ Pauli strings. However, with $\mathcal{O}(M)$ Pauli strings, the number of coefficients to tune is still $N^M$. As such, procedures like those found in \cite{wilde2022scalably, franca2022efficient, gu2022practical} will only consider $M = 2$, corresponding to quadratic Hamiltonians, with a geometric locality constraint such as nearest neighbour coupling only.
With NODEs, we it is possible to go to $\mathcal{O}(M>2)$ Hamiltonians with a curriculum learning strategy as discussed in the main paper. Here, we briefly describe a second curriculum that can be applied in case of convergence issues for Hamiltonians with higher complexity. To that end, let $\theta_m$ denote the set of trainable parameters associated to an $\mathcal{O}(m < M)$ Hamiltonian. For example, with 
\begin{equation}
\begin{split}
    H_A &= \sum_{j} c_j^x X_{j} X_{j + 1} + c_j^y Y_j Y_{J + 1} + c_j^z Z_j Z_{j + 1} \\
    &+ \sum_k d_k^x X_k X_{k + 1} X_{k + 2} + d_k^y Y_k Y_{k + 1} Y_{k + 2} +  d_k^z Z_k Z_{k + 1} Z_{k + 2}\\,
\end{split}
\end{equation}
we have $\theta_2 = (c^x_j, c^y_j, c^z_j)$ and $\theta_3 = (d^x_k, d^y_k,d^z_k$). This prescription separates parameters to be trained in batches according to the polynomial order of their accompanying Pauli string. Having separated different orders, we apply the following curriculum:
\subsection*{Curriculum 2}
\begin{itemize}

    \item [i] Switch off all $\theta_m$ except the smallest order. i.e. for all $n > m_{min}$, $\theta_n = 0$.
    \item [ii] Apply \textbf{Curriculum 1} with $\theta_m$ and a NN via Fig.\ref{fig:architectures_c}. \ref{fig:node_architecture}.
    \item [iii] Freeze $\theta_m$, and allow $\theta_{m + 1} \neq 0$
    \item [ii] Apply \textbf{Curriculum 1} with $\theta_{m + 2}$ and a NN via Fig.\ref{fig:architectures_c}. \ref{fig:node_architecture}.
    \item [iv] Repeat steps [ii], [iii] until the desired polynomial order, $\theta_M$ is reached.
\end{itemize}
The result of applying this curriculum is that the measurement outcomes become more sensitive to variation of sub-components of the Hamiltonian, mitigating any possible issues that can arise with vanishing gradients. In the main text, this curriculum was not applied as our test set of Hamiltonians was convergent using the simpler curriculum in Sec. \ref{sec:curriculum_learning}. However, we believe this more elaborate curriculum is worth mentioning here because future works in this direction will need to apply it.
\end{document}